\documentclass[onecolumn,superscriptaddress,
amsmath,amssymb, aps, pre,]{revtex4-1}
\usepackage{dcolumn}
\usepackage{bm}
\usepackage{graphicx}
\usepackage{comment}
\usepackage[caption=false]{subfig}
\usepackage{color}

\newcommand{\bea}{\begin{eqnarray}}

\newcommand{\eea}{\end{eqnarray}}

\begin{document}

\preprint{}
\title{Active particle in one dimension subjected to resetting with memory}
\author{Denis Boyer}
\affiliation{Instituto de F\'\i sica, Universidad Nacional Aut\'onoma de M\'exico, 
Ciudad de M\'exico 04510, M\'exico}
\author{Satya N. Majumdar}
\affiliation{LPTMS, CNRS, Univ.  Paris-Sud,  Universit\'e Paris-Saclay,  91405 Orsay,  France}

\begin{abstract}

The study of diffusion with preferential returns to places visited in 
the past has attracted an increased attention in recent years. In these 
highly non-Markov processes, a standard diffusive particle 
intermittently resets at a given rate to previously visited positions. 
At each reset, a position to be revisited is randomly chosen with a 
probability proportional to the accumulated amount of time spent by the 
particle at that position. These preferential revisits typically 
generate a very slow diffusion, logarithmic in time, but still with a 
Gaussian position distribution at late times. Here we consider an active 
version of this model, where between resets the particle is 
self-propelled with constant speed and switches direction in one 
dimension according to a telegraphic noise. Hence there are two sources 
of non-Markovianity in the problem. We exactly derive the position 
distribution in Fourier space, as well as the variance of the position 
at all times. The crossover from the short-time ballistic regime, 
dominated by activity, to the large-time anomalous logarithmic growth 
induced by memory is studied. We also analytically derive a large 
deviation principle for the position, which exhibits a logarithmic 
time-scaling instead of the usual algebraic form. Interestingly, at large 
distances, the large deviations become independent of time and match the 
non-equilibrium steady state of a particle under resetting to its 
starting position only.
\end{abstract}

\maketitle

\section{Introduction}

Consider a single particle diffusing on a line with diffusion constant 
$D$. In Ref.~\cite{BS_2014}, this diffusive dynamics was studied in the 
presence of an additional resetting move that is history dependent. In 
this model, in addition to diffusion, the particle also undergoes 
resetting with rate $r$ to a previously visited position according to 
the following stochastic rule. At any given time $t$, the particle 
chooses at random ({\it i.e.}, with probability density $1/t$) a 
preceding time $t'\in[0,t)$, and resets to the position at which it was 
located at $t'$. Thus the resetting move tends to dynamically localize 
the particle near the positions that are most often visited, as the 
probability to choose a particular site for revisit is proportional to 
the accumulated occupation time at that site. The position distribution 
of the particle was found to approach a Gaussian form at late times $t$, 
with the variance growing in a slow, anomalous way as $(2D/r)\, \ln 
(rt)$ asymptotically~\cite{BS_2014}. This very slow dynamics emerges 
from resetting induced memory effects: the particle becomes sluggish and 
\lq\lq reluctant" to move away from its familiar territory that it has 
already visited. Moreover, the position distribution is always 
time-dependent and does not approach a stationary state at late times, 
in contrast to diffusion with stochastic resetting to a single site 
\cite{EM_2011,EMS_review}.  This simple model was able to fit quantitatively several statistical properties of the movements of capuchin monkeys in the wild \cite{BS_2014}, bringing further evidence that many animal species use sophisticated cognitive skills to explore their environment and that memory should be incorporated in biological random walk models \cite{BDF_2008,F_2008,VM_2009,F_2013,MFM_2014,FC_2021,F_2023}.

Various other generalisations of this simple, exactly 
solvable model have been studied in recent 
years~\cite{BR_2014,BP_2016,BEM_2017,FBGM_2017,BFGM_2019}. 
Moreover, central and local limit 
theorems have been rigorously established for an extended class of 
memory walks of this type (including cases with fading memory), the 
proofs being based on a mapping onto weighted random recursive 
trees~\cite{MU_2019}. The large deviations of these walks were the focus of Ref. \cite{BM_2023}. A quenched large deviation principle was proven when the sequence of resetting times was given and acting as a disordered environment \cite{BM_2023}.

\begin{figure}[b]
\includegraphics[width=0.6\textwidth]{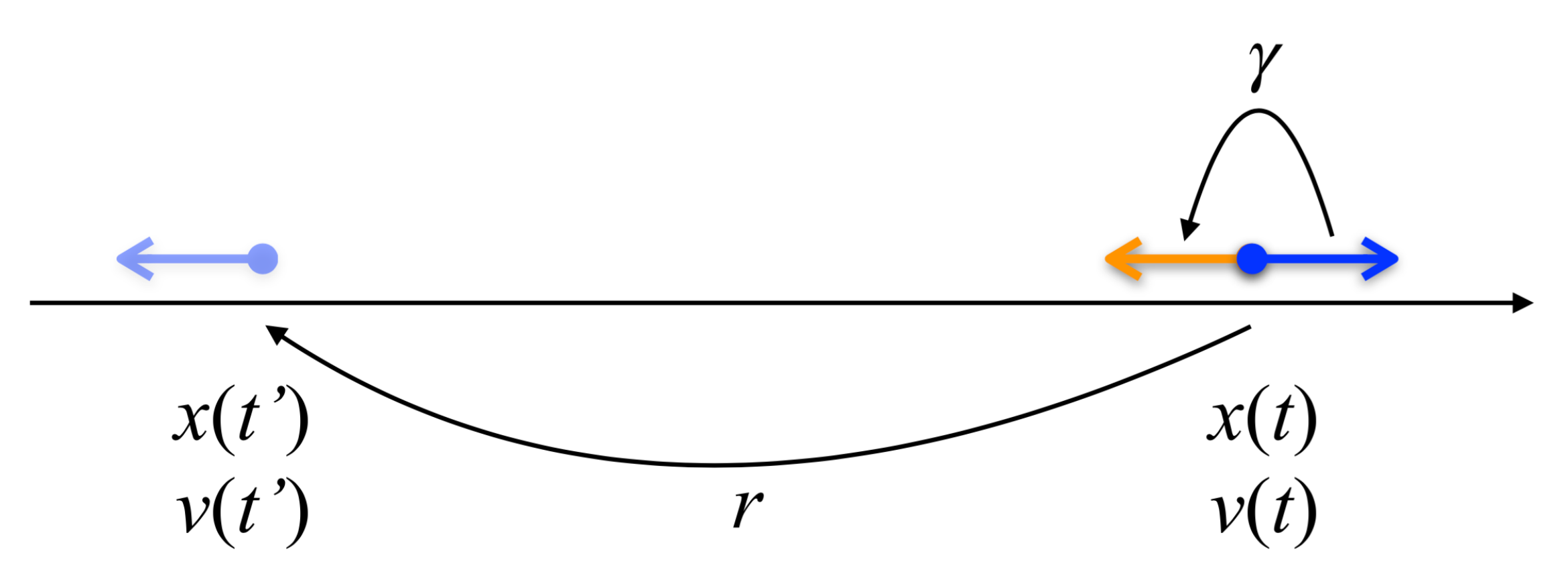}
\caption{ An active particle (dark blue arrow) moves at constant speed $v_0$ on the line and  stochastically switches direction at rate $\gamma$. Besides, the particle can relocate instantaneously to a previously visited location with rate $r$. In this case, a time $t'\in [0,t)$ is chosen at random uniformly in the past and the particle adopts the position and velocity it had at that time (light blue arrow). }
\label{fig.diag}
\end{figure}

In this paper, we consider an active version of this model where the resetting dynamics is similar as above, 
except that between resets the particle undergoes a self-propelled active dynamics in one spatial dimension with velocities $\pm v_0$, instead
of ordinary diffusion. Correlated random walks have been widely used to model the directional persistence which characterizes the movements of many animals and cells (see, e.g., \cite{CPB_2008} for a review).
The state of the particle at time $t$ is now specified by two degrees of freedom: the
position and the velocity.
More precisely, in a small time $dt$, with probability $1-r\, dt$, 
the position $x(t)$ of the particle gets
incremented by $v_0\, \sigma(t)\, dt$, where $\sigma(t)=\pm 1$ is a telegraphic noise that switches
between $+1$ and $-1$ with a constant rate $\gamma$. The driving noise has a two-time correlation
function that decays exponentially with the time difference,
\begin{equation}
 \langle v_0\, \sigma(t)\, v_0\, \sigma(t')\rangle= v_0^2\, e^{-2\,\gamma\,|t-t'|}\, .
\label{noise_auto.1}
\end{equation}
With the complementary probability $r\, dt$, the particle undergoes resetting. If this happens, the particle chooses a previous time interval $[t',t'+dt']$ with uniform probability $dt'/t$, where $0\le t'\le t$.  It then changes the
current position $x$ to the position $x'$ it occupied at time $t'$, and also takes the velocity it had at $t'$. Thus the transition
from the state $(x, \sigma, t)$ to $(x', \sigma', t')$ occurs with probability $r\, dt\, dt'/t$ under this resetting protocol. Figure \ref{fig.diag} illustrates these dynamical rules.

In this model, there are clearly two sources of non-Markovianity. The first coming from the driving noise which has
a memory encoded in the autocorrelation function above (which for finite $\gamma$ is not delta correlated
as in the white noise case). Whereas the complete state of the particle $(x(t),\sigma(t))$ has a Markovian evolution when the motion is driven by the telegraphic noise, the marginal process $x(t)$ is non-Markovian. The second source of non-Markovianity comes from the resetting moves which depends
on the past history of the trajectory. In this case, even the process $(x(t),\sigma(t))$ is non-Markovian.
Thus the model is characterized by two time scales:
{\it (i)} $t_1^*\sim 1/\gamma$ that denotes the typical time between two consecutive switches of the driving noise
and {\it (ii)} $t_2^*\sim 1/r$ denoting the typical time between two consecutive resetting events. The interplay
between these two time-scales leads to a position distribution $p_r(x,t)$ that is rather rich
and nontrivial, as we will illustrate below. We note that a model with the same active dynamics on a
line, but subjected to a resetting only to the fixed initial position (with velocity randomized after each reset), 
was solved exactly in Ref.~\cite{EM_2018}.
The resetting protocol in our model here is thus quite different from that of Ref.~\cite{EM_2018}. Here,
under resetting, both the position and velocity of the particle get reset to the values they had at a previous time, chosen uniformly at random. Consequently, the positions that have been often visited are more likely to be chosen as a resetting point in the future. Actually, under these rules, the probability that the particle chooses a given visited location $x'$ is proportional to the total time it spent at that location.

The rest of the paper is organized as follows.
In Section II we recall known results on the position distribution of a particle driven by a telegraphic noise only ($r=0$), and then summarize our main results in the presence of memory ($r>0$).
In Section III, we present the derivation
of the exact Fourier transform $q_r(k,t)$ of the position distribution in the presence of resetting, given by
Eq. (\ref{prkt.1}) below. In Section IV, we derive the exact formula for the variance given by Eq. (\ref{var_r.1}). In Section V, we derive the large deviation form, which is summarized in  Eq. (\ref{largedev_r.1}).
Finally, we conclude and give some remarks in Section VI.
Appendix A provides a brief summary 
of the {\em quenched} large deviations principle derived in Ref.~\cite{BM_2023} by a very different probabilistic approach, along with a comparison
with our large deviation results, which correspond to the annealed case. In Appendix B we give technical details on the small-$k$ expansion of the position distribution up to order 2, from which the variance is obtained.

\section{Summary of known results and our results}

In the absence of resetting $r=0$, the position distribution of a single particle driven by a telegraphic noise has been studied extensively in the past,
going back more than 100 years~\cite{Furth_1920,Taylor_1920}. It has
repeatedly resurfaced in many different contexts, such as 
the relativistic chessboard model of Feynman~\cite{FH_1965}, the persistent random walk model
of Kac~\cite{Kac_1974,MW_1992,Weiss_2002}, in quantum optics and chemical physics~\cite{HJ_1995}, in semi-flexible
polymer chains in one dimension~\cite{DC_2002,SS_2002}, and  
more recently in the context of active 
matter~\cite{TC_2008,CT_2015,Angelani_2015,
Bechinger16,Malakar_2018,DM_2018,Dhar_2019,GM_2019,MLMS_2021}. 
Assuming that the particle starts at the origin $x=0$ with equally likely velocity $\pm v_0$, let
$p_0(x,t)$ denote the position distribution at time $t$, where the subscript $0$ refers
to zero resetting or $r=0$. Its Fourier transform has an exact expression at all $t$~\cite{Weiss_2002,Malakar_2018}
\begin{equation}
q_0(k,t) = \int_{-\infty}^{\infty} p_0(x,t)\, e^{i k x}\, dx= e^{-\gamma t}\, \left[
\cosh\left( \sqrt{\gamma^2- v_0^2 k^2}\, t\right)+ \frac{\gamma}{\gamma^2-v_0^2 k^2}\,
\sinh\left( \sqrt{\gamma^2- v_0^2 k^2}\, t\right)\right]\, .
\label{p0kt_r0.1}
\end{equation}
By taking derivatives with respect to $k$ at $k=0$, one can calculate all the moments at all $t$. For example,
the mean is zero by symmetry and the variance $V_0(t)= - \partial_k^2 q_0(k,t)\Big|_{k=0}$ is given by
\begin{equation}
V_0(t)= \frac{v_0^2}{\gamma^2}\left[ \gamma\, t- \frac{1}{2}\left(1- e^{-2 \gamma\, t}\right)\right]\, .
\label{var_r0.1}
\end{equation}
Since $r=0$, we have only one time-scale $t_1^*\sim 1/\gamma$ in the problem. Indeed, the
variance grows differently as a function of $t$, as one crosses this time scale,
\begin{eqnarray}
V_0(t) \simeq \begin{cases}
v_0^2\, t^2 \quad\,\quad {\rm for} \quad t\ll 1/\gamma \\
\\
2\, D_{\rm eff}\, t \quad\, {\rm for} \quad t\gg 1/\gamma\, ,
\end{cases}
\label{var_r0.2}
\end{eqnarray}
where $D_{\rm eff}= v_0^2/(2\gamma)$ is the effective diffusion constant. This indicates that for $t\ll 1/\gamma$, where the noise $\sigma$
is yet to flip, the particle moves ballistically, while for $t\gg 1/\gamma$, the particle undergoes
diffusion with an effective diffusion constant $D_{\rm eff}$.
Interestingly, in this $r=0$ case, the Fourier transform in Eq. (\ref{p0kt_r0.1}) can be exactly inverted to give ~\cite{Weiss_2002,DC_2002,Malakar_2018}
\begin{equation}
p_0(x,t)= \frac{1}{2}\, e^{-\gamma\, t}\left[ \delta(x-v_0 t) + \delta(x+v_0 t)+
\frac{\gamma}{2v_0}\, \left(I_0(\rho)+ \frac{\gamma t}{\rho}\, I_1(\rho)\right)\, \theta(v_0t-|x|)\right]\, ,
\quad {\rm with}\quad \rho= \frac{\gamma}{v_0} \, \sqrt{v_0^2 t^2-x^2}\, ,
\label{p0xt_r0.1}
\end{equation}
where $I_0(z)$ and $I_1(z)$ are modified Bessel functions of the first kind of order $0$ and $1$ respectively.
Thus, for finite $t$, the distribution $p_0(x,t)$ is supported over a finite interval $x\in [-v_0\, t, v_0\, t]$,
with two symmetrical delta peaks located at the two edges and a central part that has a Gaussian shape near $x=0$.
As $t\to \infty$, the amplitude of the delta peaks at the two edges vanishes exponentially fast, and for 
$t\gg 1/\gamma$, the typical fluctuations of $O(\sqrt{t})$ are distributed via the Gaussian form 
\begin{equation}
p_0(x,t) \simeq \frac{1}{\sqrt{4\pi D_{\rm eff}\, t}}\, e^{- \frac{x^2}{4  D_{\rm eff}\, t}} \, , \quad {\rm with}\quad
D_{\rm eff}= \frac{v_0^2}{2\gamma}\, . 
\label{gauss_r0.1}
\end{equation}
However, the large atypical fluctuations, say of $O(t)$, are not described by the central Gaussian form.
In fact, from the exact distribution in Eq. (\ref{p0xt_r0.1}), one finds that both the typical
and the atypical fluctuations, for large $t$, are captured by a single large deviation form \cite{DMS_2021}
\begin{equation}
p_0(x,t)\sim e^{- \gamma\, t\, \Phi_0\left(\frac{x}{v_0 t}\right)}\, , 
\label{large_dev_r0.1}
\end{equation}
where the rate function $\Phi_0(z)$, supported over $z\in [-1,1]$, is given explicitly by
\begin{equation}
\Phi_0(z)= 1-\sqrt{1-z^2}\, \quad\, -1\le z\le 1 \, .  
\label{phi0z.1}
\end{equation}
In the limit $z\to 0$, {\it i.e.}, when $|x|\ll v_0 t$, the rate function becomes quadratic $\Phi_0(z)\approx z^2/2$. Substituting this
quadratic form in Eq. (\ref{large_dev_r0.1}), one recovers the Gaussian shape in Eq. (\ref{gauss_r0.1}).

In this paper, we provide an exact solution of the position distribution 
$p_r(x,t)$ when the resetting with rate $r\ge 0$
is switched on. Let us first summarize our main results. We assume, as in the $r=0$ case, that the
particle starts at the origin with equally likely velocities $\pm v_0$. We then compute exactly, for all $t$, the Fourier transform
of the position distribution $q_r(k,t)= \int_{-\infty}^{\infty} p_r(x,t)\, e^{i k x}\, dx$ and show
that it is given by the formula
\begin{equation}
q_r(k,t)= \frac{1}{2}\left[\left(1- \frac{\gamma}{\sqrt{\gamma^2-v_0^2\, k^2}}\right) \, f_{\lambda_1(k)}(t) +
\left(1+ \frac{\gamma}{\sqrt{\gamma^2-v_0^2\, k^2}}\right)\, f_{\lambda_2(k)}(t)\right]\, , 
\label{prkt.1}
\end{equation}
where
\begin{equation}
\lambda_1(k)= \gamma+\sqrt{\gamma^2-v_0^2\, k^2} \, \quad {\rm and}\quad \lambda_2(k)= \gamma-\sqrt{\gamma^2-v_0^2\, k^2}\, . 
\label{lambda12_def}
\end{equation}
The function $f_{\lambda}(t)$ is given explicitly by
\begin{equation}
f_{\lambda}(t)=  M\left( \frac{\lambda}{r+\lambda}, 1, - (r+\lambda)\, t\right) ,
\label{flt.1}
\end{equation}
where $M(a,b,z)$ is the confluent hypergeometric function of the first kind
(Kummer's function) that has a simple
power series expansion around $z=0$~\cite{AS_book}:
\begin{equation}
M(a,b,z)= 1 + \frac{a}{b}\, z+ \frac{a(a+1)}{b(b+1)}\, 
\frac{z^2}{2!}+ \frac{a(a+1)(a+2)}{b(b+1)(b+2)}\, \frac{z^3}{3!}+ \ldots\, .
\label{M_def.1}
\end{equation}
For $r=0$, using the identity $M(1,1,z)=e^z$, it is easy to see that Eq. (\ref{prkt.1}) reduces to
the expression in Eq. (\ref{p0kt_r0.1}). 

\begin{figure}
\includegraphics[width=0.7\textwidth]{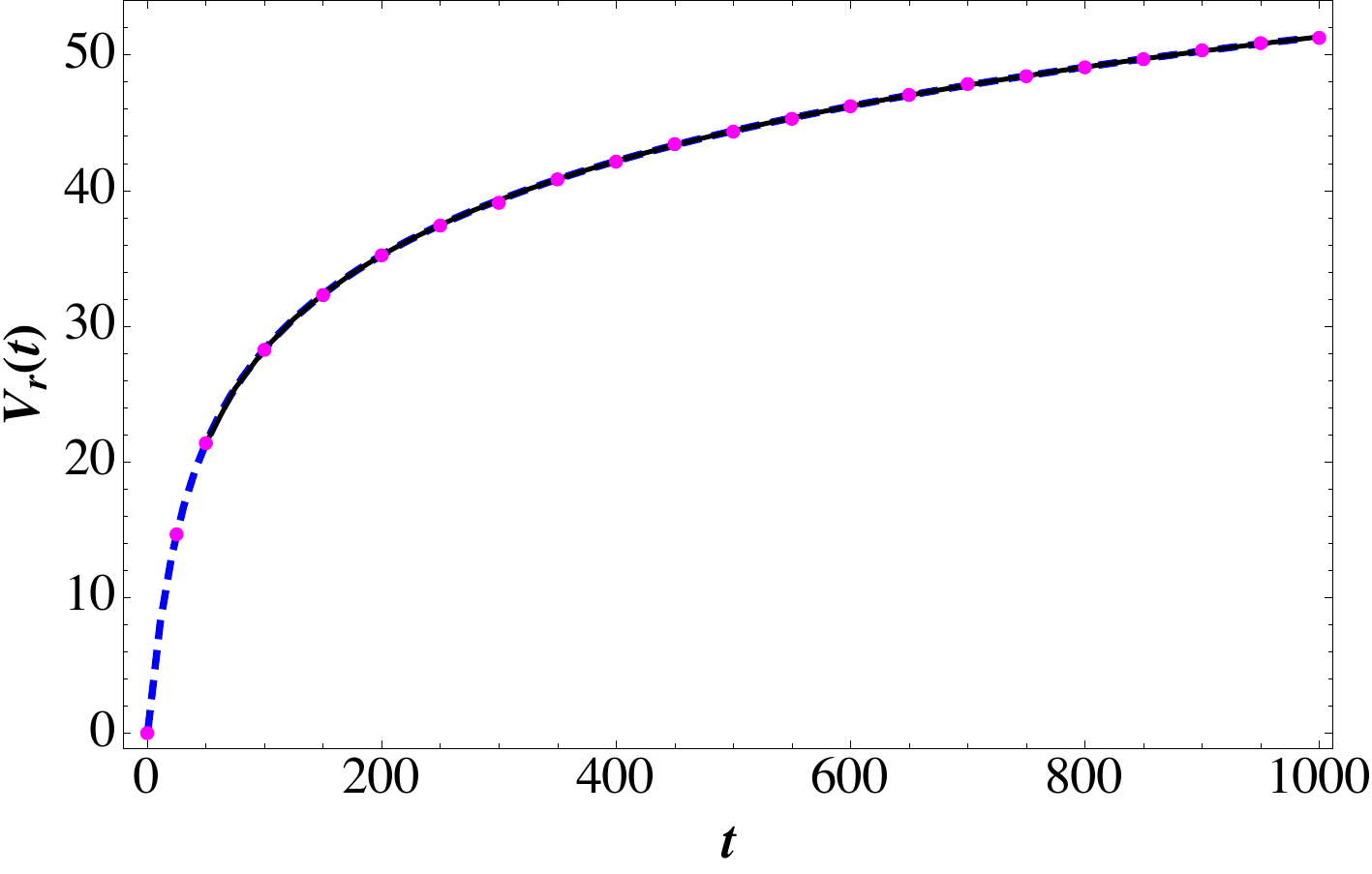}
\caption{The variance $V_r(t)$ of the position $x(t)$ as a 
function of $t$, for $v_0=1$, $\gamma=1$ and $r=0.1$.  The (blue) dashed line corresponds to the exact result in Eq. (\ref{var_r.1}) and the (black) solid line to the long time behavior in Eq. (\ref{var_asymp}). The dots are results of Monte Carlo simulations employing the Gillespie algorithm  \cite{G_1976}.}
\label{fig.var_r}
\end{figure}

This exact Fourier transform in Eq. (\ref{prkt.1}), valid for all $t$, gives
access to the moments of $x(t)$. For example, while the mean is zero at all times by symmetry, the variance $V_r(t)$ is given
by the explicit expression, valid for all $t$,
\begin{equation}
V_r(t)= \frac{v_0^2}{2\gamma^2}\, \left[ M\left(\frac{2\gamma}{r+2\gamma}, 1, - (r+2\gamma)\, t\right)-1
+ \frac{2\gamma}{r}\, \left(\ln (rt)+\gamma_E + \Gamma(0, r\, t)\right)\right]\, ,
\label{var_r.1}
\end{equation}
where $\Gamma(0,z)= \int_z^{\infty} \frac{e^{-x}}{x}\, dx$ (for $z>0$) and $\gamma_E=0.5772156649\ldots$ is the
Euler's constant. A plot of $V_r(t)$ in Eq. (\ref{var_r.1}) as a function of $t$ is provided in Fig. \ref{fig.var_r}, for
fixed $v_0$, $\gamma$ and $r$, showing very good agreement with numerical simulations. Again, for $r=0$, it is easy to check that we recover the result in Eq. (\ref{var_r0.1}). For a nonzero $r$, we have now two times scales $t_1^*\sim 1/\gamma$ and $t_2^*\sim 1/r$.
Suppose we have a small resetting rate $r$ so that $t_2^*\gg t_1^*$. The
exact variance in Eq. (\ref{var_r.1}) 
exhibits three different growth regimes:
{\it (i)} a short time regime $t\ll t_1^*=1/\gamma$ where the variance grows quadratically with $t$, {\it (ii)} an intermediate
regime $1/\gamma\ll t\ll 1/r$ where the variance grows diffusively, and finally {\it (iii)} a late time regime
$t\gg 1/r$ where the variance grows extremely slowly as $\sim \ln (rt)$. The precise asymptotic behaviors of 
the variance are given by
\begin{eqnarray}
V_r(t) \simeq \begin{cases}
v_0^2\left[ t^2 - \frac{(12\gamma+7 r)}{18}\, t^3 + O(t^4)\right]\, \quad\quad\quad\quad\quad\quad {\rm as} \quad t\to 0 \\
\\
\frac{v_0^2}{r \, \gamma}\, \left[ \ln (rt) + \left(\gamma_E - \frac{r}{2\gamma}\right) + O\left(t^{-\frac{2\gamma}{r+2\gamma}}\right)
\right] \quad {\rm as} \quad t\to \infty\, .
\end{cases}
\label{var_asymp}
\end{eqnarray}
The latter expression is also represented in Fig. \ref{fig.var_r}.

Unfortunately, unlike in the $r=0$ case, we were not able to invert the 
the Fourier transform $q_r(k,t)$
in Eq. (\ref{prkt.1}). Nevertheless the exact Fourier 
transform allows us to extract the large deviation
form of $p_r(x,t)$ for large $t$. We find that it satisfies an anomalous large deviation form
\begin{equation}
p_r(x,t) \sim (rt)^{-\Psi_r\left(\frac{\sqrt{\gamma\, r}}{v_0}\, 
\frac{x}{\ln (rt)}\right)}= 
e^{- \ln (rt)\, \Psi_r\left(\frac{\sqrt{\gamma\, r}}{v_0}\, \frac{x}{\ln (rt)}\right)}\, ,
\label{largedev_r.1}
\end{equation}
where the rate function $\Psi_r(z)$ is symmetric $\Psi_r(z)=\Psi_r(-z)$ and can be analytically computed.
It has the following asymptotic behaviors
\begin{eqnarray}
\Psi_r(z)\simeq \begin{cases}
\frac{z^2}{2} \quad\quad\quad\quad\,\,\, {\rm as} \quad z\to 0 \\
\\
\sqrt{2+ \frac{r}{\gamma}}\, |z| \quad {\rm as}\quad z\to \pm \infty\, .
\end{cases}
\label{Psi_asymp}
\end{eqnarray}
Substituting the quadratic behavior for small $z$ in Eq. (\ref{largedev_r.1}), one finds that
the typical fluctuations of $O\left(\sqrt{\ln (rt)}\right)$ are described by a Gaussian form
\begin{equation}
p_r(x,t) \sim \exp\left[- \frac{\gamma\, r\, x^2}{2\, v_0^2\, \ln (rt)}\right]\, .  
\label{typical.1}
\end{equation}
This result is consistent with late time behavior of the variance in the second line
of Eq. (\ref{var_asymp}). We note that anomalous large deviation behavior
of the type in Eq. (\ref{largedev_r.1}), where $\ln(rt)$ plays the role of 
an effective
time, has been found also in quite a few unrelated systems such as in persistence problems of Gaussian 
stationary processes~\cite{MB_1998,BMS_2013}, in the statistics of the
zeroes of random Kac-type polynomials~\cite{GS_2007,GS_2008,PS_2018}, in models of rainfall records~\cite{MVK_2019}
and in certain predator-prey systems~\cite{EMS_2022}. 
We thus present here analytical expressions where
such an anomalous large deviation form holds with a logarithmic scaling 
instead of the standard $t$. 
Recently, an anomalous large deviations principle was also 
proven rigorously by very different methods for a general class of memory models similar to the one considered here, and where the 
time evolution was governed by the law $[\ln(rt)]^{\alpha}$ 
with $\alpha>0$ \cite{BM_2023}. The large deviations principle of  \cite{BM_2023} corresponds to the quenched case, i.e., when the resetting times are fixed and not averaged over like here.
In Appendix \ref{appA} we show that our analytical expression for $\Psi_r(z)$ [see Eq. (\ref{max_action.1}) below] can be re-derived by using the general relation of \cite{BM_2023}.

\vspace{0.5cm}

\section{The exact position distribution}

We consider an active self-propelled particle on the line whose position, between resetting events, evolves
via $dx/dt= v_0\sigma(t)$, where $\sigma(t)=\pm 1$ is the telegraphic noise that
switches between the two states with rate $\gamma$. The state of the particle
at time $t$ is specified by two degrees of freedom $(x, \sigma)$, namely the 
position and the velocity (in units of $v_0$). 
Let us recall the resetting dynamics. At time $t$, with probability $r dt'/t$ the particle chooses any previous 
time $t'\le t$ and resets, {\it i.e.} the position and velocity get reset to the values taken at $t'$.
Let $p_r(x,\sigma,t)$ denotes the probability density
that the particle is at position $x$ with velocity $\sigma$ at time $t$. 
To take into account the memory effect, we also need to define the
two-point function $p_r(x,\sigma,t\,; \, x', \sigma',t')$, which denotes the joint probability density for the particle 
to be at $(x', \sigma')$ at time $t'$ and 
and at $(x, \sigma)$ at $t$, with $t'\le t$. Clearly, if we integrate the two-point functions 
over $(x',\sigma')$ (or alternately
over $(x,\sigma)$), we recover the
marginal one-point probability density
\begin{equation}
\sum_{\sigma'=\pm 1}\int_{-\infty}^{\infty} p_r(x,\sigma,t\, ;\, x', \sigma', t')\, dx'= 
p_r(x,\sigma, t)\, , \quad {\rm and} \,\, {\rm similarly}\quad \sum_{\sigma=\pm 1}\int_{-\infty}^{\infty}
p_r(x, \sigma, t\, ;\, x', \sigma', t')\, dx= p_r(x',\sigma', t')\, .
\label{marg.1}
\end{equation}
With these ingredients at hand, we can now write down a Fokker-Planck equation for the evolution of the one-point
position distributions $p_r(x,\sigma=1, t)\equiv p_r^{+}(x,t)$ and $p_r(x,\sigma=-1,t)\equiv p_r^{-}(x,t)$ as follows.
\begin{eqnarray}
\partial_t p_r^+(x,t)&= & -v_0\, \partial_x p_r^+(x,t) - \gamma\, p_r^+(x, t) + \gamma\, p_r^{-}(x, t)
-r\, p_r^+(x,t) + \frac{r}{t} \, \int_0^t dt' 
\sum_{\sigma=\pm 1}\int_{-\infty}^{\infty} dx'\, p_r(x', \sigma, t;\, x, +1, t')\, \label{pr_plus.1} \\
\partial_t p_r^-(x,t)&= & v_0\, \partial_x p_r^-(x,t) - \gamma\, p_r^-(x, t) + \gamma\, p_r^{+}(x, t)
-r\, p_r^-(x,t) + \frac{r}{t} \, \int_0^t dt' 
\sum_{\sigma=\pm 1}\int_{-\infty}^{\infty} dx'\, p_r(x', \sigma, t;\, x, -1, t')\, . \label{pr_minus.1}
\end{eqnarray}
The first three terms on the right hand side (r.h.s.) of this pair of equations describe
the standard evolution under the self-propelled dynamics. The third term
in the first line (respectively the second line) describes the loss of probability density from position 
$(x, +1)$ (respectively $(x,-1)$) at time $t$ due to resetting to other coordinates.
The last term on the first line (respectively the second line) describes the gain in the probability density 
at $(x,+1)$ (respectively at $(x,-1)$) due to resetting
from other positions occupied at time $t$ just before resetting and labeled by $(x', \sigma)$. If the particle has to reset to $(x,+1)$ (respectively
$(x,-1)$) from $(x',\sigma)$ 
at time $t$, it must have been
at $(x, +1)$ (respectively $(x,-1)$) at a previous time $t'\le t$ and the probability density of this event 
is simply the
two-point function $p_r(x', \sigma, t; x, +1, t')$. Finally, we need to integrate over all $(x', \sigma)$ from
which the particle may arrive at $(x,+1)$ (respectively $(x,-1)$) by resetting. The reason for the solvability 
for the one-point
position distribution in this model can then traced back to the second relation in Eq. (\ref{marg.1}), which
allows us to write a closed pair of equations for the one-point function only, without involving higher-point functions,
\begin{eqnarray}
\partial_t p_r^+(x,t)&= & -v_0\, \partial_x p_r^+(x,t) - \gamma\, p_r^+(x, t) + \gamma\, p_r^{-}(x, t)
-r\, p_r^+(x,t) + \frac{r}{t} \, \int_0^t dt' p_r^+(x, t') \label{pr_plus.2} \\
\partial_t p_r^-(x,t)&= & v_0\, \partial_x p_r^-(x,t) - \gamma\, p_r^-(x, t) + \gamma\, p_r^{+}(x, t)
-r\, p_r^-(x,t) + \frac{r}{t} \, \int_0^t dt' p_r^{-}(x,t')\, . \label{pr_minus.2}
\end{eqnarray}
These two equations start from the initial conditions
\begin{equation}
p_r^+(x,0)= p_r^{-}(x,0)= \frac{1}{2} \delta(x)\, , 
\label{ic.1}
\end{equation}
which corresponds to a particle starting at the origin with equally likely velocities
$\pm v_0$. The boundary conditions are $p_r^{\pm}(x\to \pm \infty, t)=0$. 
The full position distribution at time $t$ is then obtained by summing the two solutions
\begin{equation}
p_r(x,t)= p_r^+(x,t) + p_r^{-}(x,t)\, .
\label{position_def.1}
\end{equation}
By integrating Eqs. (\ref{pr_plus.2}) and (\ref{pr_minus.2}) over $x$  and adding, one can easily check
that the total probability $\int_{-\infty}^{\infty} p_r(x,t)\, dx=1$ is conserved at all times.
Evidently, for $r=0$, these equations reduce to the standard pair of Fokker-Planck equations
studied extensively in the literature~\cite{Weiss_2002,Malakar_2018}.

To solve the pair of coupled partial differential equations (\ref{pr_plus.2}) and (\ref{pr_minus.2}),
it turns out to be convenient to work in the Fourier space. We define the pair of Fourier transforms
\begin{equation}
q_r^{\pm}(k,t)= \int_{-\infty}^{\infty} p_r^{\pm}(x,t)\, e^{i k x}\, dx\, .
\label{FT_def.1}
\end{equation}
Taking Fourier transforms of Eqs. (\ref{pr_plus.2}) and (\ref{pr_minus.2}) we obtain 
\begin{eqnarray}
\partial_t q_r^+(k,t)&= & (i\, v_0\, k -\gamma-r) \, q_r^+(k,t)+ \gamma\, q_r^{-}(k,t)
+\frac{r}{t}\, \int_0^t q_r^+(k,t')\, dt' \label{pk_plus.1} \\
\partial_t q_r^{-}(k,t)&= & -(i\, v_0\, k +\gamma+r) \, q_r^-(k,t)+ \gamma\, q_r^{+}(k,t)
+\frac{r}{t}\, \int_0^t q_r^-(k,t')\, dt' \, . \label{pk_minus.1} 
\end{eqnarray}
The initial conditions (\ref{ic.1}) translate, in the Fourier space, to
\begin{equation}
q_r^{+}(k,0)= q_r^{-}(k,0)= \frac{1}{2}\, .
\label{ic.2}
\end{equation}

The equations (\ref{pk_plus.1}) and (\ref{pk_minus.1}) are still coupled and time-dependent. To make
further progress we make the following ansatz
\begin{equation}
q_r^{\pm}(k,t)= f(t,k)\, u_{\pm} (k)\, .
\label{ansatz.1}
\end{equation}
Note that this ansatz is not really a separation of variables 
since $f(t,k)$ may depend on both $t$ and $k$.
It just assumes that both $q_r^{+}(k,t)$ and $q_r^{+}(k,t)$ have the same time dependence
through the common shared factor $f(t,k)$. The goal would be to find these functions $f(t,k)$ and
$u_{\pm}(k)$. Substituting (\ref{ansatz.1}) into the pair of equations (\ref{pk_plus.1})
and (\ref{pk_minus.1}) and dividing both sides by $f(t,k)\, u_{\pm}(k)$ gives
\begin{eqnarray}
\frac{1}{f(t,k)}\, \partial_t f(t,k) - \frac{r}{t\, f(t,k)}\, \int_0^{t} f(t',k)\, dt' &= & 
(i\,v_0\, k-\gamma-r) +
\gamma\, \frac{u_{-}(k)}{u_+(k)} \label{fk_plus.1} \\
\frac{1}{f(t,k)}\, \partial_t f(t,k) - \frac{r}{t\, f(t,k)}\, \int_0^{t} f(t',k)\, dt' &= &
-(i\, v_0\, k+ \gamma+r) + \gamma\, \frac{u_{+}(k)}{u_{-}(k)}\, . \label{fk_minus.1}
\end{eqnarray} 
It thus follows that the right hand sides of these two equations cannot be a function of $t$,
and depend only on $k$. It then follows immediately that the function $f(t,k)$ must satisfy an eigenvalue-like equation
\begin{equation}
\frac{1}{f(t,k)}\, \partial_t f(t,k) - \frac{r}{t\, f(t,k)}\, \int_0^{t} f(t',k)\, dt'= -r-\lambda(k) \, ,
\label{eigen_ft.1}
\end{equation}
where we defined the eigenvalue $\lambda(k)$ with a shift by $r$ for convenience. 
From Eqs. (\ref{fk_plus.1} and (\ref{fk_minus.1}), it then follows that
the eigenvalue $\lambda(k)$ must satisfy the pair of equations
\begin{eqnarray}
(i\,v_0\, k-\gamma) +
\gamma\, \frac{u_{-}(k)}{u_+(k)} &= & -\lambda(k)\, \label{u_plus.1} \\
-(i\, v_0\, k+ \gamma) + \gamma\, \frac{u_{+}(k)}{u_{-}(k)}&=& -\lambda(k) \, . \label{u_minus.1}
\end{eqnarray}
To satisfy both (\ref{u_plus.1}) and (\ref{u_minus.1}), the eigenvlaue $\lambda(k)$ must satisfy
\begin{equation}
\left(\lambda(k)+ i\, v_0\, k -\gamma\right)\,\left(\lambda(k)-i\, v_0\ k-\gamma\right)-\gamma^2=0\, ,
\label{eigen.2}
\end{equation}
which has two roots
\begin{equation}
\lambda_{1}(k)= \gamma+ \sqrt{\gamma^2-v_0^2\, k^2} \quad {\rm and}\quad \lambda_2(k)= \gamma- 
\sqrt{\gamma^2-v_0^2\, k^2}\, .
\label{two_roots.1}
\end{equation}
In addition, for each of these eigenvalues, it follows from Eqs. (\ref{u_plus.1}) and (\ref{u_minus.1})
that $u_{\pm}(k)$ must also satisfy the relation
\begin{equation}
\frac{u_+(k)}{u_{-}(k)}= - \frac{\lambda(k)-i\, v_0\, k-\gamma}{\gamma}
=-\frac{\gamma}{\lambda(k)+i\, v_0\, k-\gamma}\, .
\label{eigen_rel.1}
\end{equation}

Given the eigenvalue $\lambda(k)$, we now need to determine the function $f(t,k)$ from Eq. (\ref{eigen_ft.1}).
To solve this equation, we first define 
\begin{equation}
F(t,k)= \int_0^t f(t', k)\, dt'\, ,
\label{FT_def.2}
\end{equation}
which then satisfies, using (\ref{eigen_ft.1}), the second order ordinary differential equation (for fixed $k$)
\begin{equation}
t\, \frac{d^2 F(t,k)}{dt^2} + (r+\lambda(k))\, t\, \frac{dF(t,k)}{dt} - r\, F(t,k)=0\, ,
\label{FT_diff.1}
\end{equation}
subject to the condition
\begin{equation}
F(0,k)=0\, 
\label{F0k.1}
\end{equation}
that follows from the definition (\ref{FT_def.2}). We now make the substitution
\begin{equation}
F(t,k)= t\, W\left(-(r+\lambda(k))\, t\right)\, 
\label{Ftk_sub.1}
\end{equation}
in Eq. (\ref{FT_diff.1}). Then it is straightforward to show that $W(z)$ satisfies the
ordinary differential equation
\begin{equation}
z\, W''(z)+ (2-z)\, W'(z)- \frac{\lambda(k)}{r+\lambda(k)}\, W(z)=0\, .
\label{Wz_diff.1}
\end{equation}
This is the standard confluent hypergeometric differential equation~\cite{AS_book} whose
general solution can be written as the linear combination of two independent solutions
\begin{equation}
W(z)= c_1\, M\left( \frac{\lambda(k)}{r+\lambda(k)}, 2, -z\right) + c_2\, 
U\left(\frac{\lambda(k)}{r+\lambda(k)}, 2, - z\right)\, ,
\label{Wz_sol.1}
\end{equation}
where $c_1$ and $c_2$ are arbitrary constants. Thus, using Eq. (\ref{Ftk_sub.1}), the general solution of 
$F(t,k)$ can be expressed as
\begin{equation} 
F(t,k)= c_1\, t\, M\left(\frac{\lambda(k)}{r+\lambda(k)}, 2, -(r+\lambda(k))\, t\right) + c_2\,t\, 
U\left(\frac{\lambda(k)}{r+\lambda(k)}, 2, - (r+\lambda(k))\, t\right)\, .
\label{Ftk_sol.1}
\end{equation}
However, this solution must satisfy the constraint $F(0,k)=0$ in (\ref{F0k.1}). Now, using
the small argument asymptotics of the two solutions~\cite{AS_book}
\begin{eqnarray}
M(a,b,z) &\to & 1\, \quad\quad\quad\,  {\rm as}\quad z\to 0 \label{Mz0.1} \\
U(a,b,z) &\to & \frac{1}{\Gamma(a)\, z} \, \quad {\rm as}\quad z\to 0 \, , 
\label{Uz0.1} 
\end{eqnarray}
one finds that
\begin{equation}
F(0,k)\to - \frac{c_2}{(r+\lambda(k)) \, \Gamma\left(\frac{\lambda(k)}{r+\lambda(k)}\right)}\, .
\label{F0.2}
\end{equation}
Since $F(0,k)=0$, we must have $c_2=0$. Hence our solution simply reads
\begin{equation}
F(t,k) = c_1\, t\, M \left(\frac{\lambda(k)}{r+\lambda(k)}, 2, -(r+\lambda(k))\, t\right)\, .
\label{Ftk_sol.2}
\end{equation} 
Taking derivative with respect to $t$ and using the definition of $M(a,b,z)$ as a power series of $z$ we  get
\begin{equation}
f_{\lambda(k)}(t,k)= \frac{dF(t,k)}{dt}= c_1\,  M \left(\frac{\lambda(k)}{r+\lambda(k)},
1, - (r+\lambda(k))\, t\right)\, ,
\label{ftk_sol.1}
\end{equation}
where we have used the subscript $\lambda(k)$ in $f(t,k)$ to display explicitly its 
dependence on the eigenvalue 
$\lambda(k)$.

Thus, given the two eigenvalues $\lambda_1(k)$ and $\lambda_2(k)$ in Eq. (\ref{two_roots.1})
and the associated time-dependent parts $f_{\lambda_1(k)}(t,k)$ and $f_{\lambda_2(k)}(t,k)$, we
can then write the complete solution (coming back to the ansatz (\ref{ansatz.1}) and using relation in Eq. (\ref{eigen_rel.1}) as
the linear combinations
\begin{eqnarray}
q_r^+(k,t) & = & u_1^+(k)\, f_{\lambda_1(k)}(t,k) + u_2^+(k)\, f_{\lambda_2(k)}(t,k) \label{qkp.2} \\
q_r^{-}(k,t) &= & - \frac{\lambda_1(k)+i\, v_0\, k-\gamma}{\gamma}\, u_1^+(k)\, f_{\lambda_1(k)}(t,k)
-\frac{\lambda_2(k)+i\, v_0\, k-\gamma}{\gamma}\, u_2^+(k)\, f_{\lambda_2(k)}(t,k)\, . \label{qkm.2}
\end{eqnarray}
Note that we can absorb the constant $c_1$ that appears in $f_{\lambda(k)}(t,k)$ in Eq. (\ref{ftk_sol.1}) in the functions
$u_1^+(k)$ and $u_2^+(k)$. In other words, we can set $c_1=1$ in the expression for $f_{\lambda(k)}(t,k)$
in Eq. (\ref{ftk_sol.1}) without any loss of generality. With this convention and using $M(a,b,0)=1$, we then obtain from Eq. (\ref{ftk_sol.1}) that
$f_{\lambda(k)}(t=0,k)=1$.
The only unknown functions $u_1^+(k)$ and $u_2^+(k)$ in Eqs. (\ref{qkp.2})
and (\ref{qkm.2}) are to be determined from the pair of initial
conditions in Eq. (\ref{ic.2}). Setting $t=0$ in Eqs. (\ref{qkp.2}) and (\ref{qkm.2}), and using 
$f_{\lambda(k)}(t=0,k)=1$, the initial condition (\ref{ic.2}) gives the
two equations
\begin{eqnarray}
u_1^+(k)+ u_2^+(k) &= & \frac{1}{2} \label{u1.1}\\
- \frac{\lambda_1(k)+i\, v_0\, k-\gamma}{\gamma}\, u_1^+(k)-\frac{\lambda_2(k)+i\, v_0\, k-
\gamma}{\gamma}\, u_2^+(k) &=& \frac{1}{2} \, . \label{u2.1}
\end{eqnarray}
Solving this pair of equations gives
\begin{eqnarray}
u_1^{+}(k)& = & -\frac{\gamma- \sqrt{\gamma^2-v_0^2\, k^2}+ i\, v_0\, k}{4\, \sqrt{\gamma^2-v_0^2\, k^2}} \label{u1_sol.1} \\
u_2^+(k) &=&  \frac{\gamma+ \sqrt{\gamma^2-v_0^2\, k^2}+ i\, v_0\, k}{4\, \sqrt{\gamma^2-v_0^2\, k^2}}\, . 
\label{u2_sol.1}
\end{eqnarray}
Substituting the expressions for $u_1^+(k)$, $u_2^+(k)$ and $f_{\lambda(k)}(t,k)$ in Eqs. (\ref{qkp.2})
and (\ref{qkm.2}) then gives us the pair of Fourier transforms $q_r^{\pm}(k,t)$ explicitly, By gathering the two terms, the Fourier transform of the full position distribution $q_r(k,t)$ is obtained as
\begin{eqnarray}
q_r(k,t)= \int_{-\infty}^{\infty} p_r(x,t)\, e^{i k x}\, dx &= & q_r^+(k,t)+q_r^{-}(k,t) \\
&= &\frac{1}{2}\left[\left(1- \frac{\gamma}{\sqrt{\gamma^2-v_0^2\, k^2}}\right) \, 
f_{{\lambda}_{1}(k)}(t) +
\left(1+ \frac{\gamma}{\sqrt{\gamma^2-v_0^2\, k^2}}\right)\, 
f_{{\lambda}_{2}(k)}(t)\right]\, , 
\label{prkt.2}
\end{eqnarray}
where $\lambda_1(k)= \gamma+\sqrt{\gamma^2-v_0^2\, k^2}$, $\lambda_2(k)=\gamma-\sqrt{\gamma^2-
v_0^2\, k^2}$ and the function $f_{\lambda(k)}(t,k)$ is given in
Eq. (\ref{ftk_sol.1}) (with $c_1=1$). This result in Eq. (\ref{prkt.2}) was announced in
Eq. (\ref{prkt.1}) in the Introduction.

For later usage, it is also useful to consider the time integrated position distribution
\begin{equation}
P_r(x,t)= \int_0^t p_r(x,t')\, dt'\, .
\label{ti_prxt.1}
\end{equation}
The Fourier transform of $P_r(x,t)$ is given by
\begin{equation}
Q_r(k,t)= \int_{-\infty}^{\infty} P_r(x,t)\, e^{i k x}\, dx=\int_0^{t} q_r(k,t')\, dt' \, . 
\label{Qrk_def.1}
\end{equation}
Using Eq. (\ref{prkt.2}), we find
\begin{equation}
Q_r(k,t)= \frac{1}{2}\left[\left(1- \frac{\gamma}{\sqrt{\gamma^2-v_0^2\, k^2}}\right) \,
F_{{\lambda}_{1}(k)}(t,k) +
\left(1+ \frac{\gamma}{\sqrt{\gamma^2-v_0^2\, k^2}}\right)\,
F_{{\lambda}_{2}(k)}(t,k)\right]\, ,
\label{Qrkt.2}
\end{equation}
where, again,  $\lambda_1(k)= \gamma+\sqrt{\gamma^2-v_0^2\, k^2}$, $\lambda_2(k)=\gamma-\sqrt{\gamma^2-
v_0^2\, k^2}$ and $F_{\lambda(k)}(t,k)$ is given in Eq. (\ref{Ftk_sol.2}) (with $c_1=1$), namely,
\begin{equation}
F_{\lambda(k)}(t,k)=  t\, M \left(\frac{\lambda(k)}{r+\lambda(k)}, 2, -(r+\lambda(k))\, t\right)\, ,
\label{Fkt_sol.2}
\end{equation}
with $M(a,b,z)$ given in Eq. (\ref{M_def.1}).

\section{Exact computation of the variance}

From the exact Fourier transform of the position distribution in Eq. (\ref{prkt.2}), one can, 
in principle, compute all the moments by making a small $k$ expansion. By symmetry,
the mean position is identically zero, $\langle x\rangle(t)=0$ at all times. Thus the variance is
given by
\begin{equation}
V_r(t)= \langle x^2\rangle(t)- [{\langle x\rangle}(t)]^2= - \frac{d^2q_r(k,t)}{dk^2}\Big|_{k=0}\, .
\label{var_r.2}
\end{equation}
To perform the small $k$ expansion of $q_r(k,t)$, it turns out to be convenient to
work with time integrated Fourier transform in Eq. (\ref{Qrkt.2}). 
For small $k$, keeping terms up to $O(k^2)$, one gets 
\begin{equation}
1- \frac{\gamma}{\sqrt{\gamma^2-v_0^2\, k^2}}\simeq - \frac{v_0^2k^2}{2\gamma^2}
\quad {\rm and}\quad 1+ \frac{\gamma}{\sqrt{\gamma^2-v_0^2\, k^2}}\simeq 2+ \frac{v_0^2k^2}{2\gamma^2}\, .
\label{amp_smallk.1}
\end{equation}
Expanding the eigenvalues in Eq. (\ref{two_roots.1}) up to $O(k^2)$ we get
\begin{equation}
\lambda_1(k)\simeq 2\gamma- \frac{v_0^2k^2}{2\gamma}\quad {\rm and}
\quad \lambda_2(k)\simeq \frac{v_0^2 k^2}{2\gamma}\, .
\label{eigen_smallk.1}
\end{equation}
Using these expansions in Eq. (\ref{Qrkt.2}), we get after some steps (see Appendix \ref{appb} for details),
\begin{equation}
Q_r(k,t) = t\, \left[1- \frac{v_0^2 k^2}{4 \gamma^2}\,\left(
-1 + M\left(\frac{2\gamma}{r+2\gamma}, 2, - (r+2\gamma)\, t\right)
-\gamma\, t\, {}_2F_2\left[\{1,1\}, \{2,3\}, -rt\right]\right)\right] + O(k^4)\, .
\label{Qrkt.4}
\end{equation}
Furthermore, it turns out that the function ${}_2F_2\left[\{1,1\}, \{2,3\}, -rt\right]$ can be
expressed in terms of elementary functions as
\begin{equation}
{}_2F_2\left[\{1,1\}, \{2,3\}, z\right]= -\frac{2}{z^2}\, \left[ e^z -1-z +\gamma_E\, z + z\, \Gamma(0,-z)
+z \ln (-z)\right]\, ,
\label{F2_iden.1}
\end{equation}
where $\Gamma(0,z)=\int_z^{\infty} e^{-x} dx/x$ for $z>0$ and $\gamma_E$ is the Euler's constant.

Deriving Eq. (\ref{Qrkt.4}) twice with respect to $k$ and setting $k=0$ gives the time integrated variance
\begin{equation}
\int_0^{t} V_r(t')\, dt'= \frac{v_0^2}{2\gamma^2}\, \left[-t +t\, M\left(\frac{2\gamma}{r+2\gamma}, 2, - (r+2\gamma)\, t\right)
+ \frac{2\gamma}{r^2}\, g(rt)\right]\, ,
\label{int_var.1}
\end{equation}
where
\begin{equation}
g(y)=1-e^{-y}+ y \ln (y) + y\, \int_y^{\infty} \frac{e^{-x}}{x}\, dx - (1-\gamma_E)\, y\, .
\label{gy_def.1}
\end{equation}
Finally, taking a derivative with respect to $t$ gives our final exact expression for the variance,
valid for all $t$ and given by Eq. (\ref{var_r.1}) in the Introduction.  In the limit $r\to 0$, it is easy to show,
using $M(1,2,z)= (e^z-1)/z$, that Eq. (\ref{var_r.1}) reduces to the known result in Eq. (\ref{var_r0.1}).
One can also derive the asymptotic behaviors of $V_r(t)$ for small
and large $t$, as shown in Eq. (\ref{var_asymp}).

\section{Large deviation form at late times}\label{sec:largedev}

While we can not invert the Fourier transform $q_r(k,t)$ in Eq. (\ref{prkt.2}) to obtain
the position distribution $p_r(x,t)$ in real space for all $t$, we show in this section that
one can make progress at large times. We show below that $p_r(x,t)$
indeed admits an anomalous large deviation form as in Eq. (\ref{largedev_r.1}). 
For large $t$, the asymptotic behavior of the Fourier transform $q_r(k,t)$ in Eq. (\ref{prkt.2}) can be derived
by using the following property of $M(a,b,z)$~\cite{AS_book}
\begin{equation}
M(a,b,-z) \simeq \frac{\Gamma(b)}{\Gamma(b-a)}\, z^{-a} \quad {\rm as}\quad z\to \infty\, .
\label{Mz_asymp.1}
\end{equation}
Using this result in Eq. (\ref{prkt.2}), one finds that for $t\gg 1/(r+2\gamma)$ (and fixed $k$) the leading term
of the Fourier transform
$q_r(k,t)$ decays as (up to a prefactor independent of time) 
\begin{equation}
q_r(k,t) \sim \left[\left(1+\frac{\lambda_2(k)}{r}\right)r\,t\right]^{-\frac{\lambda_2(k)}{r+\lambda_2(k)}}
\sim (r\,t)^{-\frac{\lambda_2(k)}{r+\lambda_2(k)}}= e^{- \frac{\lambda_2(k)}{r+\lambda_2(k)}\, \ln (rt)}\, .
\label{leading_qrkt.1}
\end{equation}
The contribution coming from $\lambda_1(k)$ is subleading since $\lambda_1(k)>\lambda_2(k)$ in
Eq. (\ref{two_roots.1}). Inverting formally the Fourier transform (\ref{leading_qrkt.1}) and using
$\lambda_2(k)= \gamma-\sqrt{\gamma^2-v_0^2\, k^2}$, one gets to leading order for large $t$
\begin{equation}
p_r(x,t) = \int_{-\infty}^{\infty} \frac{dk}{2\pi}\, e^{-i\, k\, x}\, q_r(k,t) \sim 
\int_{-\infty}^{\infty} \frac{dk}{2\pi}\, \exp\left[-i\, k\, x - \frac{\gamma-\sqrt{\gamma^2-v_0^2\, k^2}}{r+
\gamma-\sqrt{\gamma^2-v_0^2\, k^2}}\, \ln (rt) \right]\, .
\label{prxt_ldv.1}
\end{equation}

\begin{figure}
\includegraphics[width=0.4\textwidth]{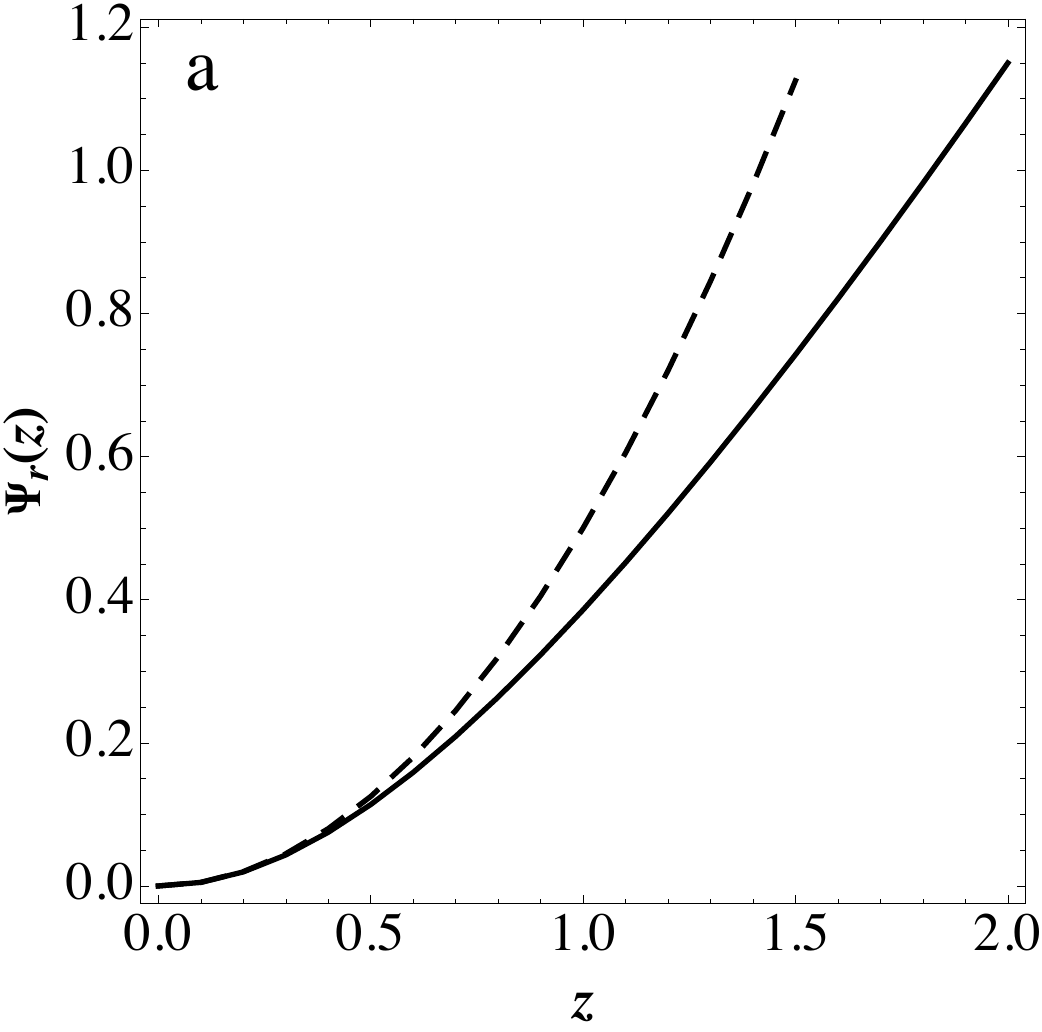}\hspace{0.5cm}
\includegraphics[width=0.41\textwidth]{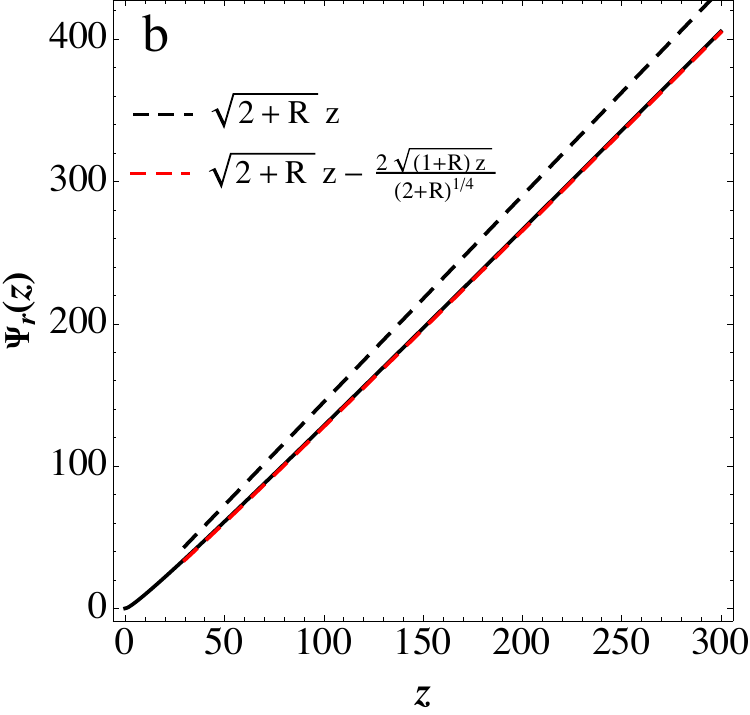}
\caption{ a) The analytical rate function $\Psi_r(z)$ 
in Eq. (\ref{max_action.1}) is plotted against $z$ (solid line)
for fixed $r=0.1$ and $\gamma=1$  (or $R=r/\gamma=0.1$),  while the quadratic behavior at small $z$ in Eq. (\ref{Psi_asymp.1}) is displayed with a dashed line. b) Same rate function at large $z$ (solid line), with the leading asymptotic form in Eq. (\ref{Psi_asymp.1}) as a black dashed line. The red dashed line represents the leading term plus the first sub-leading correction.}
\label{fig.ratef_r}
\end{figure}

It is now convenient to make a change of variable 
$i k\, v_0/\sqrt{\gamma\, r}=q$ (Wick's rotation) and rewrite the 
$k$-integral as a Bromwich integral in the complex $q$ plane
\begin{equation}
p_r(x,t)\sim \frac{\sqrt{\gamma\, r}}{v_0}\, \int_{\Gamma} \frac{dq}{2\pi i} \exp \left[-\ln (rt)\, \left( 1+q\, z-
\frac{R}{1+R-\sqrt{1+R\, q^2}}\right)\right]\, \quad {\rm with}\quad R= \frac{r}{\gamma}\, .
\label{wick_rot.1}
\end{equation}
The contour $\Gamma$ runs vertically in the complex $q$ plane without a real shift and we have defined
\begin{equation}
z= \frac{\sqrt{\gamma\, r}\, x}{v_0\, \ln (rt)}\, .
\label{z_def.1}
\end{equation}
Let us first remark that $p_r(x,t)$ in Eq. (\ref{prxt_ldv.1}) is clearly symmetric in $x$, or equivalently 
in Eq. (\ref{wick_rot.1}) as a function
of the scaled variable $z$. Hence, without any loss of generality, we will just consider
the case $z\ge 0$.

Now, for large $t$, one can
evaluate the integral in Eq. (\ref{wick_rot.1}) by the saddle point method. This gives the desired
large deviation form
\begin{equation}
p_r(x,t) \sim \exp\left[- \ln (rt)\, \Psi_r(z)\right]\, \quad {\rm with}\quad z= \frac{\sqrt{\gamma\, r}\, x}{v_0\, \ln (rt)}\, ,
\label{ldv_final.1}
\end{equation}
where the rate function $\Psi_r(z)$ is symmetric around $z=0$, {\it i.e.}, $\Psi_r(-z)=\Psi_r(z)$.
For $z\ge 0$, the rate function $\Psi_r(z)$ is obtained from the following maximization ({\it i.e.}, the saddle point analysis
of Eq. (\ref{wick_rot.1}))
\begin{equation}
\Psi_r(z)= \sup\limits_{0\le q\le q_{\max}} \, \left[ 1+ q\, z- \frac{R}{1+R-\sqrt{1+R\, q^2}}\right]\, .
\label{max_action.1}
\end{equation}
Note that in the maximization over $q$ above, the range of $q$ is limited over $q\in [0, q_{\max}]$
where $q_{\max}$ is given by 
\begin{equation}
1+R-\sqrt{1+R\, q_{\max}^2}=0 \quad {\rm or}\quad q_{\max}=\sqrt{R+2}\, . 
\label{qmax_def.1}
\end{equation}

To compute the rate function $\Psi_r(z)$, we proceed as follows. Let us first define
\begin{equation}
S(q,z)= 1+ q\, z- \frac{R}{1+R-\sqrt{1+R\, q^2}}\, .
\label{Sqz_def}
\end{equation}
Deriving with respect to $q$ and setting it to zero, {\it i.e.}, $\partial_q S(q,z)=0$  gives $q=q^*(z)$ where
\begin{equation}
z= \frac{q^*\,  R^2}{ \sqrt{1+R\, (q^*)^2} \, \left[1+R - \sqrt{1+R\, (q^*)^2}\right]^2}\, .
\label{zq*.1}
\end{equation}
Now evaluating the action $S(q,z)$ at $q=q^*$ gives the rate function
\begin{equation}
\Psi_r(z)= S(q^*(z), z)= 1- \frac{R}{1+R- \sqrt{1+R\, (q^*(z))^2}}+ q^*(z)\, z
\label{action.2}
\end{equation}
where $z$ depends on $q^*$ through Eq. (\ref{zq*.1}).
One needs to eliminate $q^*$ from Eqs. (\ref{action.2}) and (\ref{zq*.1}) to express $\Psi_r(z)$
as a function of $z$. This can be done using parametric plot in Mathematica, that lets us plot
$\Psi_r(z)$ as a function of $z$ in Fig. \ref{fig.ratef_r}{\color{blue}a}. Note that we have only
plotted $\Psi_r(z)$ for $z\ge 0$. For the negative argument, one simply uses the symmetry
$\Psi_r(-z)=\Psi_r(z)$.

One can also derive the asymptotic behaviors of $\Psi_r(z)$ as $z\to 0$ and $z\to \infty$.
To derive these behaviors, let us first express $\Psi_r(z)$ in a convenient form.
Indeed, by taking a total derivative of $\Psi_r(z)$ in Eq. (\ref{action.2}), we get
\begin{equation}
\frac{d\Psi_r(z)}{dz}= q^*(z) + \partial_q S(q,z)\Big|_{q=q^*(z)}\, \frac{dq^*(z)}{dz}= q^*(z)\, ,
\label{der_ratef.1}
\end{equation}
where the last equality holds because the second term on the r.h.s. vanishes, as $\partial_q S(q,z)=0$ at the saddle point $q=q^*(z)$. 
Now integrating Eq. (\ref{der_ratef.1}) with respect to $z$ gives the exact identity (assuming $z\ge 0$)
\begin{equation}
\Psi_r(z) = \int_0^z q^*(z')\, dz' \, ,
\label{Psirz_exact.1}
\end{equation}
where we used the fact that $\Psi_r(z=0)=0$. Thus, we need to first solve $q^*(z)$ implicitly as a function of $z$
from Eq. (\ref{zq*.1}) and then use it in Eq. (\ref{Psirz_exact.1}) to derive $\Psi_r(z)$. This gives
us access to the asymptotic behaviors of $\Psi_r(z)$. For example, when $z\to 0$, we get from Eq. (\ref{zq*.1})
that $q^*(z)\simeq z$. Hence, from Eq. (\ref{Psirz_exact.1}), we get $\Psi_r(z) \simeq z^2/2$.
In contrast, when $z\to \infty$, we expect $q^*\to q_{\max}=\sqrt{2+R}$ defined in Eq. (\ref{qmax_def.1}).
Hence, from Eq. (\ref{Psirz_exact.1}), we obtain a linear growth $\Psi_r(z)= \sqrt{R+2}\, z$ for large $z$.
By symmetry, it behaves as $\Psi_r(z)= \sqrt{R+2}\, |z|$ as $z\to -\infty$. 
In summary, using $R=r/\gamma$, the two asymptotic limits are given by
\begin{eqnarray}
\Psi_r(z)\simeq \begin{cases}
\frac{z^2}{2} \quad\quad\quad\quad\,\,\, {\rm as} \quad z\to 0 \\
\\
\sqrt{2+ \frac{r}{\gamma}}\, |z| \quad {\rm as}\quad z\to \pm \infty\, .
\end{cases}
\label{Psi_asymp.1}
\end{eqnarray}
The quadratic form and the crossover to linear behavior as $z$ increases can be seen in Fig. \ref{fig.ratef_r}a. Figure \ref{fig.ratef_r}b shows that the rate function at large $z$ tends to the linear form in Eq. (\ref{Psi_asymp.1}). We have also represented this asymptotic behavior plus the first sub-leading correction, of order $|z|^{1/2}$, giving a very good agreement with the actual $\Psi_r(z)$. We have not attempted to perform a comparison of the large deviation function with numerical simulations because of the very slow convergence toward the large $t$ asymptotic limit. As already noted in Ref. \cite{BS_2014}, the convergence to the Gaussian central part is extremely slow, as the leading corrections are of the order of $1/\ln(t)$ and still not negligible at the times that can be accessed numerically.

The two asymptotic limits of the rate function $\Psi_r(z)$ have interesting physical
implications. For $z\to 0$, substititung the quadratic behavior of $\Psi_r(z)$ in the
large deviation form in Eq. (\ref{ldv_final.1}), one gets
\begin{equation}
p_r(x,t) \sim \exp\left[- \frac{\gamma\, r\, x^2}{2\,v_0^2\, \ln (rt) } \right]\, \quad {\rm for} \quad  |x|\ll \ln (rt)\, ,
\label{smallz.1}
\end{equation} 
which shows that the typical fluctuations of $O\left(\sqrt{\ln (rt)}\right)$ around the mean are Gaussian
distributed. Moreover, the variance $v_0^2 \ln (rt)/(\gamma\, r)$ matches perfectly with
the leading asymptotic growth of the variance in Eq. (\ref{var_asymp}). 
Furthermore, using $D_{\rm eff}= v_0^2/(2\gamma)$, one recovers, at late times,
the Gaussian behavior with variance $(2 D_{\rm eff}/r)\, \ln (rt)$ as
in ordinary diffusion with the same memory-dependent resetting~\cite{BS_2014,BEM_2017}.
This is expected since at very late times the active dynamics is known to effectively behave
as a standard diffusion with an effective diffusion constant $D_{\rm eff}= v_0^2/(2\gamma)$.

In the opposite limt $|z|\to \infty$, we get by substituing the linear behavior (\ref{Psi_asymp.1})
into the large deviation form (\ref{ldv_final.1})
\begin{equation}
p_r(x,t) \sim \exp\left[- \frac{\sqrt{r(r+2\gamma)}}{v_0}\,  |x|\right] \quad {\rm for} \quad |x|\gg \ln (rt)\, .
\label{largez.1}
\end{equation}
This result may, at first glance, look surprising because it says that very far away from the center, {\it i.e.},
for $|x|\gg \ln (rt)$, the distribution actually becomes time-independent at late times. However,
this is what one would expect in hindsight. At late times, the particle is essentially concentrated 
in a core region around the origin of width $\sqrt{\ln (rt)}$. Seen from very far away from the core,
the dynamics essentially reduces to an active particle that is reset effectively to the origin.
This later problem was studied in Ref.~\cite{EM_2018} and the authors demonstrated that the
position distribution approaches a stationary form at late times 
which happens to exactly coincide with Eq. (\ref{largez.1}). Thus it is quite natural
that $p_r(x,t)$ in our problem, for $|x|\gg \ln (rt)$, becomes stationary as shown by Eq. (\ref{largez.1}). In the Appendix \ref{appA}, we show that the results of this Section are consistent with the recent findings of Ref. \cite{BM_2023}.

\section{Summary and Conclusion}

We have studied an active diffusion model with memory in one spatial 
dimension, in which a self-propelled particle moves at constant speed 
$v_0$ and switches direction stochastically at rate $\gamma$. Memory 
effects are implemented through a protocol where the particle interrupts 
its active motion and resets at rate $r$ to positions and velocities 
that were occupied at 
previous times. Namely, during a time interval $[t,t+dt]$, with 
probability $rdt$ the particle chooses a time $t'$ in the past, 
uniformly distributed in $[0,t]$, and takes the position and velocity it 
had at time $t'$, before continuing its active motion from there. This 
model extends previous studies that considered random walks or Brownian 
particles with similar resetting protocols 
\cite{BS_2014,BR_2014,BP_2016,BEM_2017}. The case studied here also 
differs significantly from the same active process in which resetting 
occurs to a unique position ({\it e.g.}, the starting position) 
\cite{EM_2018}, due to the fact that it lacks a steady state. The memory 
rule makes the particle return more often to positions frequently 
occupied (typically near its starting position), generating a diffusive 
process in which the variance of the position grows very slowly with 
time, but without being arrested asymptotically.

We have calculated exactly the distribution of the position of the 
particle in Fourier space, extending the results of Ref. \cite{BEM_2017} 
for Brownian motion. While we were not able to invert this Fourier
transform, we used 
it to calculate exactly the full time dependence of the variance of the 
position. The interplay between activity and memory leads to a variance 
growing as $t^2$ at short times and as $\ln(rt)$ as large times, this 
latter regime being a fingerprint of preferential visit models. The 
knowledge of the position distribution, whose central part becomes 
Gaussian at large times, allowed us to derive a large deviation 
principle and to study the rate function explicitly for the first time 
for this type of processes. At distances much larger than a typical 
scale $\ln(rt)$, the distribution of the position becomes exponential, 
independent of time and coincides with the non-equilibrium steady state 
produced by resetting an active particle to its starting position only.

This problem could be generalized in several directions in the future, for instance, to other protocols such as those involving periodic resetting. The results recently derived in Ref. \cite{BM_2023} for large deviations could be applied to this case and many others. Systems of many independent particles  become correlated over time when subject to simultaneous resetting \cite{BHMS_2023,BKMS_2023}. How correlations are built though resetting in preferential revisit models is also a question that deserves further study.

\acknowledgements SNM wants to thank the hospitality of KITP, Santa Barabra where this work 
initiated during the workshop ``Out-of-equilibrium Dynamics and Quantum Information of 
Many-body Systems with Long-range Interactions" (October, 2023), supported in part by the 
National Science Foundation under Grant No. NSF PHY-1748958. DB acknowledges support from Conacyt (Mexico) grant Ciencia de Frontera 2019/10872.

\appendix

\section{A quenched large deviation principle for Brownian and active particles with memory}\label{appA}

We show that our results on large deviations, in particular the 
expression (\ref{max_action.1}) of the rate function, can be derived
by an alternative method 
using a recent theorem of Boci and Mailler valid for a more general class of 
processes with preferential revisits \cite{BM_2023}. We first recall 
briefly their main results. Let $z(t)$ be a Markov process and 
$\boldsymbol{L}=\{L_i\}_{i\in \mathbb{N}}$ an infinite set of 
independent and identically distributed
(i.i.d.) positive random 
variables each distributed via $\phi(L)$. The $L_i$'s represent the time 
intervals between successive resetting events.
During each such interval the position of the walker $x(t)$ evolves freely 
according to the process $z(t)$. When the $n$-th 
resetting occurs (at time $t_n=\sum_{i=1}^n L_i$), the walker randomly 
chooses a time in the past $t'\in[0,t_n]$ with a normalized probability density 
$\mu(t')/\int_0^{t_n}dt\,\mu(t)$ where $\mu$ is a given function, and relocates to the position it occupied at that time, {\it 
i.e.}, $x(t_n)=x(t')$. From there, $x$ continues to evolve as $z$ until 
the next resetting at time $t_{n+1}$. While we took a uniform kernel 
($\mu(t')=1$), Boci and Mailler considered a more general form
\begin{equation}
\mu(t')=\frac{\alpha}{t'}(\ln t')^{\alpha-1}e^{\beta(\ln t')^{\alpha}},
\end{equation}
which reduces to the uniform case when $\alpha=\beta=1$. 

The next goal is to find a large deviation principle for
the position $x(t)$, given that a large deviation principle
exists for the underlying free Markov process $z(t)$ between
two resettings. The prescriptions of Boci and Mailler for
computing this large deviation principle proceeds via
three steps (here we simplify the notations of Ref.~\cite{BM_2023}
to adapt to the language of physicists).

\begin{itemize}

\item
One first assumes that the free Markov process $z(t)$ admits a large 
deviation principle, {\it i.e.}, there exists a function 
$\Lambda(\zeta)$ such that
\begin{equation}\label{LambdaZ}
\lim_{t\to\infty}\frac{1}{t}\ln\langle e^{\zeta z(t)}\rangle=\Lambda(\zeta),
\end{equation}
for all $\zeta$ and with $z(t=0)$ fixed. This simply means that to
leading order for large $t$, 
\begin{equation}
\label{LambdaZ.1}
\langle e^{\zeta z(t)}\rangle \sim \exp\left[ t\, \Lambda(\zeta)\right]\, .
\end{equation} 

\item The next step is to compute the following generating function $g(\xi)$
for all $\xi$
\begin{equation}
\label{g_xi.1}
g(\xi)\equiv \frac{\langle e^{\xi L}-1-\xi L\rangle}{\xi\, \langle L\rangle}
= \frac{\int_0^{\infty} \left(e^{\xi L}-1-\xi\, L\right)\, 
\phi(L)\, dL}{\xi\, \int_0^{\infty} L\, \phi(L)\, dL}\, ,
\end{equation}
where we recall that $\phi(L)$ is the distribution of the interval $L$ between
two successive resets.

\item Knowing the two functions
$\Lambda(\zeta)$ and $g(\xi)$ from above, the final step
consists in computing the following Legendre transform
\begin{equation}\label{BM3}
\Psi(w)=\sup_{y}\{w\, y-g\left(\Lambda(y)\right)\}\, .
\end{equation}

\end{itemize}

Once $\Psi(w)$ is computed, the Boci-Mailler theorem states 
that under rather mild conditions the position $x(t)$ satisfies 
a {\em quenched} large deviation principle, {\it i.e.}, 
conditioned on the $L_i$'s. For all $w>0$, this principle can be written as
\begin{equation}\label{BM1}
\lim_{t\to\infty} \frac{\ln\left[ {\rm Prob}(|x(t)|\ge w s(t)|
\boldsymbol{L})\right]}{s(t)}=
-\inf_{y\ge w}\Psi(y)=-\Psi(w),
\end{equation}
where the last equality assumes that $\Psi(y)$ is a monotonously 
increasing function of $y$. The scaling factor $s(t)$ is given by
\begin{equation}\label{BM2}
s(t)=(\ln t)^{\alpha}\quad{\rm if}\ \beta\neq 0 \, .
\end{equation}
In the language of physicists, the statement in Eq. (\ref{BM1}) simply
means that to leading order in large $t$, the cumulative position
distribution behaves as
\begin{equation}
{\rm Prob}\left(|x(t)|\ge X\right)\sim \exp\left[- (\ln t)^{\alpha}\,
\Psi\left(\frac{X}{(\ln t)^{\alpha}}\right)\right]\, ,
\label{ldv.1}
\end{equation}
where the rate function $\Psi(w)$ is computed from Eq. (\ref{BM3}). For
the uniform memory kernel where $\alpha=1$ and $\beta=1$, Eq. (\ref{ldv.1})  
takes precisely our large deviation form in Eq. (\ref{ldv_final.1}).

The results in Eqs. (\ref{ldv.1}) and (\ref{BM3}) 
were derived in Ref.~\cite{BM_2023} by a very different probabilistic 
method, namely, via a 
mapping of the memory induced resetting random walk to a weighted random recursive tree. 
This is a quenched principle in the sense that the averages are not 
taken over $\boldsymbol{L}$. The authors of Ref~\cite{BM_2023} 
nevertheless believe that the 
same principle should hold in the more difficult annealed case, provided 
that $\phi(L)$ decays to 0 fast enough at large $L$.
Below, we apply this general prescription
to compute the large deviation principle 
in two simple cases. In both cases, we assume 
\begin{equation}
\phi(L)= r\, e^{-r\, L} \quad {\rm for}\quad L\ge 0\, ,
\label{phiL.1}
\end{equation}
where $r$ represents the resetting rate.

\vskip 0.3cm

\noindent {\bf When the underlying free process $z(t)$ is a 
standard Brownian motion in one dimension.} We first compute $\Lambda(\zeta)$ in
Eq. (\ref{LambdaZ}). This is easy since
\begin{equation}
\langle e^{\zeta\, z(t)}\rangle= \int_{-\infty}^{\infty}
\frac{dz}{\sqrt{4\,\pi\, D\, t}}\,
 e^{\zeta\, z-z^2/{4Dt}}= e^{D\,t\, \zeta^2}\, ,
\label{zt_Brown.1}
\end{equation}
implying from Eq. (\ref{LambdaZ}) that
\begin{equation}
\Lambda(\zeta)= D\, \zeta^2\, .
\label{lzeta_br.1}
\end{equation}
Next, substituting $\phi(L)$ from Eq. (\ref{phiL.1}) in (\ref{g_xi.1}), it 
is easy to see that 
\begin{eqnarray}
g(\xi)= \begin{cases}
\frac{\xi}{r-\xi} \quad\quad\quad\,\, {\rm if} \quad \xi<r \\
\\
\infty \quad\quad\quad\quad {\rm if}\quad \xi\ge r .
\end{cases}
\label{LambdaStoch}
\end{eqnarray}
We use Eq. (\ref{BM3}) with the  functions $\Lambda(\zeta)$ and $g(\xi)$ explicitly given by Eqs. (\ref{lzeta_br.1}) and (\ref{LambdaStoch}), respectively, and obtain for the Brownian 
motion with memory-induced resetting
\begin{equation}\label{LambdaBM}
\Psi_{\rm BM}(w)=\sup_{0<y<\sqrt{\frac{r}{D}}}\left\{w\, y-
\frac{y^2}{\frac{r}{D}-y^2}\right\} \quad {\rm for}\quad w\ge 0\, .
\end{equation}
The rate function $\Psi_{\rm BM}(w)$ is symmetric in $w$, hence
it suffices just to consider $w\ge 0$.
We note that the function inside the supremum in (\ref{LambdaBM})
diverges at $y=\sqrt{r/D}$, thus making the allowed
interval for $y$ bounded. 
This is similar to the situation we encountered 
in Eq. (\ref{max_action.1}) before. It is a consequence of 
Eq. (\ref{LambdaStoch}) and of the fact that $\phi(L)$ does not 
decay faster than an exponential at large $L$. 

The above result for the Brownian case can also be derived from our 
active particle 
calculation by setting $r\ll \gamma$, {\it i.e.}, 
by expanding Eq. (\ref{max_action.1}) at first order in $R$. 
Then, recalling that $D_{eff}=v_0^2/(2\gamma)$ and 
$z=\sqrt{r/(2D_{eff})} (x/\ln t)$ in the notation of 
Section \ref{sec:largedev}, we make the change of variables 
$x/\ln t\to w$ and $\sqrt{r/(2D_{eff})}q\to y$ in 
Eq. (\ref{max_action.1}), and actually 
recover Eq. (\ref{LambdaBM}) with $D=D_{eff}$.

Although the maximization of Eq. (\ref{LambdaBM}) cannot be carried out 
explicitly, one can study the behavior of $\Psi_{\rm BM}(w)$ 
at small and large $w$, in the same way as in Section \ref{sec:largedev}. 
It is easy to show that $\Psi_{\rm BM}(w)$ has the following
asymptotic behaviors
\begin{eqnarray}
\Psi_{\rm BM}(w) \simeq \begin{cases}
\frac{r}{4D}\, w^2 \quad\quad {\rm as}\quad |w|\to 0 \\
\\
\sqrt{\frac{r}{D}}\, |w| \quad\,\, {\rm as}\quad |w|\to \infty
\end{cases}
\label{Psi_BM_asymp.1}
\end{eqnarray}
Consequently, Eq. (\ref{ldv.1}) gives
\begin{eqnarray}
{\rm Prob}(|x(t)|>X)\propto \begin{cases}
\exp\left(-\frac{X^2}{\frac{4D}{r}(\ln t)^{\alpha}}\right) 
\quad\,\,\,\, {\rm as} \quad X\ll (\ln t)^{\alpha} \\
\\
\exp\left(-\sqrt{\frac{r}{D}}\, X\right)\, \quad\quad\,\, {\rm as}\quad X\gg 
(\ln t)^{\alpha}  .
\end{cases}
\label{LambdaBMasympt}
\end{eqnarray}
This result agrees with Eq. (\ref{smallz.1})
with $D$ in (\ref{LambdaBMasympt})
replaced by $D_{\rm eff}= v_0^2/(2\gamma)$
and $\alpha=1$, while the large deviations are again independent of 
time at large $X$ and coincide with the non-equilibrium steady state of 
a Brownian particle under stochastic resetting to the origin only \cite{EM_2011}.

\vskip 0.3cm

\noindent {\bf When the underlying process $z(t)$ corresponds
to a free run-and-tumble particle in one dimension.}
Here $z(t)$ is the position of the 
 active particle driven by the telegraphic noise. 
Although this process is non-Markov, we may still try to 
apply the general
prescriptions in Eqs. (\ref{LambdaZ.1})-(\ref{BM1}) 
to this case \cite{note}. To compute $\Lambda(\zeta)$ for the active particle, 
we start from the general relation 
\begin{equation}
\label{RTP_A1}
\langle e^{\zeta\, z(t)}\rangle=\int_{-\infty}^{\infty}dz\ e^{\zeta\, z}\,
P(z,t)\sim e^{\Lambda(\zeta)\,t},
\end{equation}
where $P(z,t)$ is the position distribution of the active particle at 
time $t$.  As the exact expression for $P(z,t)$ is not so simple (see Eq. (\ref{p0xt_r0.1})), we rather take the Laplace transform of Eq. (\ref{RTP_A1}) with
respect to $t$ and get
\begin{equation}\label{laplace}
\int_{-\infty}^{\infty}dz\ e^{\zeta\, z}\,\widetilde{P}(z,s)
\sim \frac{1}{s-\Lambda(\zeta)},\quad {\rm with}\ s>\Lambda(\zeta),
\end{equation}
where $\widetilde{P}(z,s)=\int_0^{\infty} dt\, e^{-st}\,P(z,t)$. 
Therefore $\Lambda(\zeta)$ is the largest real pole of the integral 
in the left hand side of Eq. (\ref{laplace}). This integral can be 
calculated from the known expression for $\widetilde{P}(z,s)$ 
(see, {\it e.g.}, \cite{EM_2018}), given explicitly by
\begin{equation}\label{rtp}
\widetilde{P}(z,s)=\frac{\lambda(s)}{2s}\ e^{-\lambda(s)|z|},
\quad{\rm with}\quad \lambda(s)=\sqrt{s(s+2\gamma)}/v_0.
\end{equation}
One then obtains,
\begin{equation}\label{laplace2}
\int_{-\infty}^{\infty}dz\ e^{\zeta\, z}\widetilde{P}(z,s)=
\frac{\lambda(s)}{2s}\left(\frac{1}{\lambda(s)-\zeta}+\frac{1}{\lambda(s)+
\zeta}\right)
\end{equation}
Assuming $\zeta>0$ without loosing generality (since the active particle process is symmetric), a pole is contained in the first term of the r.h.s. of Eq. (\ref{laplace2}), 
\begin{equation}
\frac{1}{\lambda(s)-\zeta}\simeq \frac{1}{\lambda'(s^*)(s-s^*)},
\end{equation}
where $s^*$ is the largest root of the equation $\lambda(s)-\zeta=0$. 
Identifying $s^*$ with $\Lambda(\zeta)$, one deduces from Eq. (\ref{rtp})
\begin{equation}
\label{RTP_Lzeta}
\Lambda(\zeta)=-\gamma+\sqrt{\gamma^2+v_0^2\zeta^2}.
\end{equation}
Inserting this expression in Eq. (\ref{BM3}) with $g(\xi)$ given by 
Eq. (\ref{LambdaStoch}), the rate function becomes
\begin{equation}\label{Lambdaactive}
\Psi_{\rm active}(w)=\sup_{0<y<y_{max}}\left\{w\, y-
\frac{\sqrt{1+\frac{v_0^2}{\gamma^2}y^2}-1}{1+R-
\sqrt{1+\frac{v_0^2}{\gamma^2}y^2}}\right\}\quad {\rm with}\quad R=\frac{r}{\gamma}.
\end{equation}
By making the same changes of variable as in the BM case above, 
we see that Eq. (\ref{Lambdaactive}) is equivalent to 
Eq. (\ref{max_action.1}).

\section{Small $k$ expansion of $Q_r(k,t)$ up to order 2}\label{appb}
Substituting the small $k$ expansions (\ref{amp_smallk.1}) and  (\ref{eigen_smallk.1})
in the expression (\ref{Qrkt.2}) of $Q_r(k,t)$, we get
\begin{equation}
Q_r(k,t) \simeq \frac{t}{2}\left[- \frac{v_0^2k^2}{2\gamma^2}\, \,
M\left(\frac{\lambda_1(k)}{r+\lambda_1(k)}, 2, -(r+\lambda_1(k))\, t\right) +
\left(2+ \frac{v_0^2 k^2}{2\gamma^2}\right)\,
M\left(\frac{\lambda_2(k)}{r+\lambda_2(k)}, 2, -(r+\lambda_2(k))\, t\right)
\right]\, .
\label{Qrkt.3}
\end{equation}

Since we are interested in the expansion only up to $O(k^2)$ for small $k$,
we can now put $k=0$ in the arguments of the function $M$ in the first term
on the r.h.s. of Eq. (\ref{Qrkt.3}), {\it i.e.},
\begin{equation}
M\left(\frac{\lambda_1(k)}{r+\lambda_1(k)}, 2, -(r+\lambda_1(k))\, t\right)
\simeq M\left(\frac{2\gamma}{r+2\gamma}, 2, - (r+2\gamma)\, t\right)\, .
\label{M1.1}
\end{equation}
For the second term on the r.h.s. of Eq. (\ref{Qrkt.3}), we need to expand
the arguments of $M$ for small $k$. Using Eq. (\ref{eigen_smallk.1}) we get
\begin{equation}
\frac{\lambda_2(k)}{r+\lambda_2(k)}\simeq \frac{v_0^2\, k^2}{2\gamma r}\quad
{\rm and}\quad r+\lambda_2(k)\simeq r+ \frac{v_0^2\, k^2}{2\gamma}\, .
\label{arg_M2.1}
\end{equation}
Thus the $M$ function in the second term on the r.h.s. in Eq. (\ref{Qrkt.3}) reduces to
\begin{equation}
M\left(\frac{\lambda_2(k)}{r+\lambda_2(k)}, 2, -(r+\lambda_2(k))\, t\right)
\simeq M\left( \frac{v_0^2 k^2}{2\gamma r}, 2, - \left(r+ \frac{v_0^2k^2}{2\gamma}\right)\, t\right)\, .
\label{M2.1}
\end{equation}

We now need to expand this function for small $k$ up to $O(k^2)$. This is a bit tricky.
We first expand the third argument of $M$ in Eq. (\ref{M2.1}) and get
\begin{equation}
M\left(\frac{\lambda_2(k)}{r+\lambda_2(k)}, 2, -(r+\lambda_2(k))\, t\right)
\simeq  M\left( \frac{v_0^2 k^2}{2\gamma r}, 2, -rt\right)
- \frac{v_0^2 k^2}{2\gamma}\, t\, \partial_z M\left(\frac{v_0^2k^2}{2\gamma r}, 2, z\right)\Big|_{z=-rt}\, .
\label{M2.2}
\end{equation}
Using the identity, $\partial_z M(a,b,z)= (a/b) M(a+1, b+1, z)$, we see that the second
term in Eq. (\ref{M2.2}) is of $O(k^4)$ for small $k$. Hence we can neglect the second term,
and up to $O(k^2)$, we get
\begin{equation}
M\left(\frac{\lambda_2(k)}{r+\lambda_2(k)}, 2, -(r+\lambda_2(k))\, t\right)
\simeq  M\left( \frac{v_0^2 k^2}{2\gamma r}, 2,-rt\right)\, .
\label{M2.3}
\end{equation}
Now, we need to expand the r.h.s. of Eq. (\ref{M2.3}) for small $k$. For this, it is useful to
use the following small $a$ expansion of $M(a,b,z)$
\begin{equation}
M(a,b,z) = 1+ \frac{a}{b}z\, {}_2F_2\left[\{1,1\},\{2,1+b\},z\right] +O(a^2)
\label{M_smalla.1}
\end{equation}
where ${}_2F_2$ is the generalized hypergeometric function. Hence, from Eq. (\ref{M2.3})
we get
\begin{equation}
M\left(\frac{\lambda_2(k)}{r+\lambda_2(k)}, 2, -(r+\lambda_2(k))\, t\right)
\simeq  1- \frac{v_0^2 k^2}{4\gamma}\, t\, {}_2F_2\left[\{1,1\},\{2,3\},-rt\right]\, .
\label{M2.4}
\end{equation}
Collecting all these expansions together into Eq. (\ref{Qrkt.3}), we finally get the small $k$ expansion of $Q_r(k,t)$ given by Eq. (\ref{Qrkt.4}) in the main text.


\begin{thebibliography}{}
\bibitem{note}
One may consider the active process as a two-dimensional Markov process $\bm{Z}(t)=(z(t),\sigma(t))$. The Boci-Mailler theorem applies with $\langle e^{\bm{\zeta}\cdot\bm{Z}(t)}\rangle$ in Eq. (\ref{LambdaZ}) with $\bm{\zeta}=(\zeta_z,\zeta_{\sigma})$. In the analysis, one then sets  $\xi_{\sigma}=0$.
\end{thebibliography}
\end{document}